\documentclass[mathleft,cite,graphics,graphicx,psfig,amssymb]{an}
\usepackage{graphicx}
\usepackage{amsfonts}
\usepackage{natbib}
\usepackage{times}
\begin{document}
\Pagespan{1}{}%
\Yearpublication{2008}%
\Yearsubmission{2008}%
\Month{9}%
\Volume{999}%
\Issue{88}%
\DOI{DOI}%

\title{New absolute magnitude calibrations for W Ursa Majoris type binaries}

\author{Z. Eker\inst{1,2}\fnmsep\thanks{Corresponding author:\email{eker@tug.tug.tubitak.gov.tr}\newline} 
\and S. Bilir\inst{3}
\and E. Yaz\inst{3}
\and O. Demircan\inst{2}
\and M. Helvac\i \inst{4}
}

\institute{T\"UB\.ITAK National Observatory, Akdeniz University Campus, 07058 Antalya, 
Turkey
\and
\c Canakkale Onsekiz Mart University, Faculty of Sciences and Arts, Ulup\i nar 
Astrophysical Observatory, 17100 \c Canakkale, Turkey
\and
Istanbul University, Faculty of Sciences, Department of Astronomy and 
Space Sciences, 34119 Istanbul, Turkey
\and
Ankara University, Faculty of Sciences, Department of Astronomy and 
Space Sciences, 06100 Ankara, Turkey}
\date{} 

\received{}
\accepted{}
\publonline{later} 

\abstract{Parallaxes of W UMa stars in the {\em Hipparcos} catalogue have been
analyzed. 31 W UMa stars, which have the most accurate parallaxes
($\sigma_{\pi}/\pi<0.15$) which are neither associated with a
photometric tertiary nor with evidence of a visual companion, were
selected for re-calibrating the Period--Luminosity--Color (PLC)
relation of W UMa stars. Using the Lutz--Kelker (LK) bias corrected (most probable)
parallaxes, periods ($0.26< P(day)< 0.87$), and colors
($0.04<(B-V)_{0}<1.28$) of the 31 selected W UMa, the PLC relation
have been revised and re-calibrated. The difference between the old
(revised but not bias corrected) and the new (LK bias corrected) 
relations are almost negligible in predicting the distances
of W UMa stars up to about 100 parsecs. But, it increases and may
become intolerable as distances of stars increase. Additionally,
using $(J–-H)_{0}$ and $(H–-K_{s})_{0}$ colors from {\em 2MASS}
(Two Micron All Sky Survey) data, a PLC relation working with
infrared data was derived. It can be used with infrared
colors in the range $-0.01<(J-H)_{0}<0.58$, and
$-0.10<(H-K_{s})_{0}<0.18$. Despite {\em 2MASS} data are single
epoch observations, which are not guaranteed at maximum brightness
of the W UMa stars, the established relation has been found
surprisingly consistent and reliable in predicting LK corrected
distances of W UMa stars.}

\maketitle

\section{Introduction}
Low-mass contact binaries, popularly known as W Ursa Majoris (W UMa) 
stars are easy to recognize by a light curve with equal (or nearly equal) 
depths of minima which are wide enough to touch one another. Even if 
the total luminosity produced solely by the more massive component, 
efficient energy distribution through a common envelope makes the
their surface brightness practically the same over the visible surface 
\citep{R85, R93}. Even though the effective temperature is the same 
on the surfaces of components, there could be different masses hidden under 
the common atmospheres. Their mass-ratios span from almost unity \citep[V753 
Mon, $q$=0.973,][]{Ruetal00} to very small values as small as $q$=0.066 
\citep[SX Crv,][]{Ruetal01}. Therefore, W UMa binaries are non-equilibrium 
systems exchanging mass and energy between the components. Hence, 
despite such an external simplicity, their internal structure is rather complex.

External simplicity and the observational properties of W UMa systems were 
the basic reasons for establishing an absolute magnitude calibration by 
 \cite{R94}. The calibration uses two observational quantities, the orbital
period and intrinsic color, e.g. $(B-V)$ or $(V-I)$. The period and the 
color are correlated through a combined effect of geometry, Kepler`s third 
law and main-sequence relationships. The calibration, which is in the form of
$M_{V}=M_{V}(\log P, B-V)$, has been widened later to include a metallicity 
term $[Fe/H]$ \citep{Ru95}. Those earlier calibrations were based on 18 
systems, mostly members of high galactic latitude open clusters and of
visual binaries, including only three nearby W UMa systems with known 
trigonometric parallaxes, 44i Boo B, VW Cep and $\epsilon$ CrA.

After {\em Hipparcos} parallaxes became available, the
Period--Luminosity--Color (PLC) relation has been re-calibrated
with 40 nearby W UMa systems in the solar neighborhood with 
parallaxes having relative errors from 2.7 to 24\%. With this new
data, the accuracy of the calibration ($\pm$0.25 mag at a predicted
$M_{V}$) was claimed: ``it is no longer limited by the parallax data
but, paradoxically, by the lack of reliable photometric data''
\citep{RD97}. However, later, \cite{Ru00, R04} argued: at earlier 
times, when limited information was available, it seemed that 
metallicity dependence did exist \citep{Ru95}. But later when 
extensive data for many globular clusters were examined, the 
metallicity term $[Fe/H]$ become obsolote \citep{Ru00}. 
At least, solar neighborhood calibration works well for contact 
systems of different metallicities at level of uncertainty $\sigma= 
\pm 0.28$ mag. Thus, there is no need to keep the metallicity term 
in PLC relation \citep{R04}. At last, after metallicity term was dropped, 
the color term was also removed, and the relation is transformed to a much simpler
form by \cite{R06}, when estimating spatial density of contact binaries based on the
{\em ASAS} survey as a necessity because the lack of color indices 
in the {\em ASAS} data. This calibration was established from 
21 systems of good {\em Hipparcos} ($\sigma_{\pi}/\pi<0.12$) data 
which were carefully chosen as being free of triple and multiple systems.

PLC relation provides absolute magnitudes of contact binaries if their de-reddened
colors and periods are known. Consequently, PLC relation has many practical 
usages. It is useful not only in calculating space velocities and space densities of 
W UMa \citep{R02, B05, G06} stars but also for confirming or disproving of contact 
binaries in stellar clusters \citep{R98, Ru00}. Moreover, it permits W UMa stars to 
be used as standard candles. In fact, through PLC relation, W UMa stars become
more likely to be used as standard candles than RR Lyr stars because 
W UMa stars are 24000 times more common than RR Lyr stars \citep{RD97}.

Unfortunately, random errors, presumably symmetric on the measured parallaxes, 
do not provide symmetric uncertainties on the computed distances. Therefore, 
a measured trigonometric parallax is very likely to be larger than the true 
parallax. The problem has already been noticed and studied by \cite{LK73}.
Assuming a uniform space distribution of stars and a  Gaussian distribution 
of observed parallaxes over a true parallax, \cite{LK73} have revealed that 
there exists a systematic error in the computed distances which depends only 
upon the ratio $\sigma_{\pi}/\pi$, where $\pi$ is the observed parallax.
However, this crucial study has not produced proper response in the literature. 
Many studies including existing PLC calibration were completed without 
even discussing the Lutz-Kelker bias. \cite{J01} has shown that only the careful studies 
which used parallaxes with $\sigma_{\pi}/\pi<0.1$ could be excused 
since the bias would be negligible. Otherwise, not taking the Lutz-Kelker bias 
into account would either alter the conclusions or invalidate them altogether 
if relative larger errors were involved.

Standard Lutz-Kelker corrections become significant if $\sigma_{\pi}/\pi>0.05$  
\citep{M05}. Therefore, Period-Luminosity relation of \cite{R06}, which relies 
on 21 W UMa stars with sufficiently accurate ($\sigma_{\pi}/\pi<0.12$) parallaxes, 
may not be excused. The classical PLC relation of \cite{RD97}, which was 
founded on parallaxes of 40 W UMa stars with relative errors from 2.7 to 24\%,
definitely needs to be re-calibrated. This study aims to improve existing PLC 
relation by not only correcting it according to Lutz-Kelker bias but 
also refining the sample of W UMa stars by eliminating the ones which have less
reliable parallaxes because of belonging to multiple systems.

Moreover, applicability of the PLC relation at infrared colors are investigated. 
Being based on similar principles, a similar relation (PLC) for cataclysmic 
variables (CV) has already been established by using $(J–-H)$ and $(H–-K_{s})$ 
colors from {\em 2MASS} (Two Micron All Sky Survey) data and proved to
be useful for estimating CV distances by \cite{Ak07}. Despite, {\em 2MASS} data 
are compiled from {\it single epoch} observations, which are not guaranteed to 
be at maxima, the relation which uses $(J–-H)$ and $(H–-K_{s})$
colors of {\em 2MASS} photometric system have been found to be as useful as 
the existing relation which involve $(B-V)$ and $(V-I)$ colors \citep{R04}. 
The advantage of this new relation is that it is less
effected by de-reddening problems.

\section{Data}

There are 751 W UMa binaries listed in the revised edition of the GCVS. 
Only 144 of them were found to have trigonometric parallaxes in the 
{\em Hipparcos} catalogue \citep{ESA97}. Nine of them could be discarded 
right away because their relative errors ($\sigma_{\pi}/\pi>1$) are intolerable. 
For the rest (135 systems), the distribution of relative parallax errors have 
been shown in Fig. 1.

\begin{figure}
\center
\resizebox{80mm}{64.2mm}{\includegraphics*{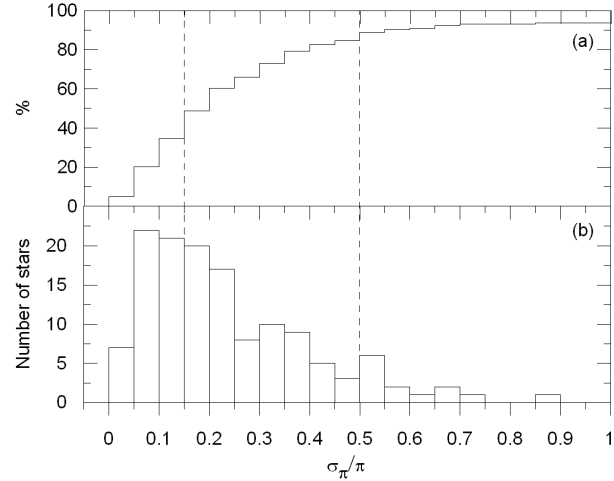}}
\caption{Relative error ($\sigma_{\pi}/\pi$) histogram of {\em Hipparcos} parallaxes 
for W UMa stars where a) shows cumulative sample and b) relative numbers. Vertical 
lines mark 50\% and 15\% relative errors.}
\end{figure}

Calibration process of PLC relation requires both accuracy of calibration
data and statistical significance. While statistical significance
requires as many as possible W UMa stars, accuracy of input
data, on the other hand, demands using only the data of W UMa stars with 
the most accurate parallaxes. Therefore, an optimized upper limit for the 
relative error of parallaxes needs to be determined. 

\subsection{Lutz-Kelker bias corrected parallaxes}

It is known that a standard Lutz-Kelker (LK) correction $<\Delta M_{V}>$, 
which corrects mean absolute magnitude of a sample, is significant if 
$\sigma_{\pi}/\pi\geq 0.05$ and increases as $\sigma_{\pi}/\pi$ increases. 
Corrections were claimed to diverge and became incomputable if 
$\sigma_{\pi}/\pi\geq0.175$ \citep{LK73, LK75, NN78, S87a, S87b, S03, B97, 
O98, M05}. Standard LK bias primarily computes an expected value of parallaxes, 
which stands for the mean of the true parallax distribution of an uniformly 
distributed sample of stars. Here, we are interested in the most probable 
value of the true parallax ($\pi_{0}$) for a single star which has an observed 
parallax ($\pi$) and associated error ($\sigma_{\pi}$). As for estimating distances, 
one can transform the posterior probability of the true parallax to that of the 
distance and use it to find the most probable distance \citep{S87a}. Considering 
that the present W UMa sample is well within the Galactic disc (TV Pic has the 
biggest $z=127$ pc, where $z$ is the distance from the galactic plane), it
can be assumed that within the limited space at the vicinity of the star, 
the distribution is uniform. Consequently, according to \cite{S87a}   
\begin{equation}
\pi_{0}=\pi(\frac{1}{2}+\frac{1}{2}\sqrt{1-16(\sigma_{\pi}/\pi)^{2}}~),
\end{equation}
is the relation between the most probable true parallax (computed) 
and the observed parallax ($\pi$) and it`s associated standard error 
($\sigma_{\pi}$). 

Although Eq. 1 is valid mathematically for $\sigma_{\pi}/\pi\leq0.25$, 
only the most probable true parallaxes of the systems with 
$\sigma_{\pi}/\pi\leq0.175$ (standard LK limit), and corresponding 
distances ($d_{0}=1/\pi_{0}$) were computed and displayed in Table 1.

\begin{table}
\setlength{\tabcolsep}{5pt}
\center 
\caption{Selection of the calibration sample from the list
of W UMa stars with parallax errors less than 17.5\%.}
{\scriptsize
\begin{tabular}{clccccccc}
\hline
     ID    &       Star &   $\pi$&  $\sigma$  &  $\sigma/\pi$ &  $\pi_{0}$ & $d_{0}$ & $d_{Hip}$& Rem. \\
           &            &      (mas) &      (mas) &      (mas) &    (mas) &       (pc) &       (pc) &        \\
\hline
         1 &    44i Boo &      78.39 &       1.03 &      0.013 &    78.34 &         13 &         13 &      1 \\
         2 &     VW Cep &      36.16 &       0.97 &      0.027 &    36.05 &         28 &         28 &      1 \\
         3 &$\epsilon$ CrA &   33.43 &       0.92 &      0.028 &    33.32 &         30 &         30 & $\surd$\\
         4 &   V972 Her &      16.25 &       0.61 &      0.038 &    16.16 &         62 &         62 & $\surd$\\
         5 &     CN Hyi &      17.22 &       0.65 &      0.038 &    17.12 &         58 &         58 &      1 \\
         6 &     AE Phe &      20.49 &       0.81 &      0.040 &    20.36 &         49 &         49 & $\surd$\\
         7 &     DO Cha &      12.56 &       0.57 &      0.045 &    12.46 &         80 &         80 & $\surd$\\
         8 &  V2082 Cyg &      11.04 &       0.56 &      0.051 &    10.92 &         92 &         91 &       1\\
         9 &      W UMa &      20.17 &       1.05 &      0.052 &    19.95 &         50 &         50 &       1\\
        10 &      S Ant &      13.30 &       0.71 &      0.053 &    13.15 &         76 &         75 & $\surd$\\
        11 &   V335 Peg &      16.26 &       0.86 &      0.053 &    16.08 &         62 &         62 &      1 \\
        12 &  V2388 Oph &      14.72 &       0.81 &      0.055 &    14.54 &         69 &         68 &      1 \\
        13 &     OU Ser &      17.31 &       0.95 &      0.055 &    17.10 &         58 &         58 & $\surd$\\
        14 &   V445 Cep &       8.95 &       0.50 &      0.056 &     8.84 &        113 &        112 &      1 \\
        15 &   V759 Cen &      15.88 &       0.93 &      0.059 &    15.66 &         64 &         63 & $\surd$\\
        16 &     AW UMa &      15.13 &       0.90 &      0.059 &    14.92 &         67 &         66 &       1\\
        17 &     GR Vir &      18.83 &       1.18 &      0.063 &    18.53 &         54 &         53 & $\surd$\\
        18 &     KR Com &      13.07 &       0.87 &      0.067 &    12.83 &         78 &         77 &       1\\
        19 &     YY Eri &      17.96 &       1.20 &      0.067 &    17.63 &         57 &         56 & $\surd$\\
        20 &     IS CMa &      10.01 &       0.71 &      0.071 &     9.80 &        102 &        100 & $\surd$ \\
        21 &     CP Hyi &       8.38 &       0.60 &      0.072 &     8.20 &        122 &        119 & $\surd$ \\
        22 &     YY CrB &      11.36 &       0.85 &      0.075 &    11.10 &         90 &         88 & $\surd$ \\
        23 &   V757 Cen &      14.18 &       1.10 &      0.078 &    13.83 &         72 &         71 & $\surd$ \\
        24 &     FX Eri &      13.67 &       1.06 &      0.078 &    13.33 &         75 &         73 & $\surd$ \\
        25 &   V566 Oph &      13.98 &       1.11 &      0.079 &    13.62 &         73 &         72 &       1 \\
        26 &     RR Cen &       9.76 &       0.85 &      0.087 &     9.45 &        106 &        103 & $\surd$ \\
        27 &     GM Dra &      10.16 &       0.88 &      0.087 &     9.84 &        102 &         98 & $\surd$ \\
        28 &   V899 Her &       8.06 &       0.77 &      0.096 &     7.75 &        129 &        124 &       1 \\
        29 &   V502 Oph &      11.84 &       1.17 &      0.099 &    11.36 &         88 &         85 &       1 \\
        30 &   V535 Ara &       8.87 &       0.90 &      0.101 &     8.49 &        118 &        113 & $\surd$ \\
        31 &     SW Lac &      12.30 &       1.26 &      0.102 &    11.76 &         85 &         81 &       1 \\
        32 &     ET Leo &      13.90 &       1.44 &      0.104 &    13.27 &         75 &         72 &       1 \\
        33 &     TY Men &       5.93 &       0.63 &      0.106 &     5.65 &        177 &        169 & $\surd$ \\
        34 &   V918 Her &       8.70 &       0.93 &      0.107 &     8.28 &        121 &        115 & $\surd$ \\
        35 &   V781 Tau &      12.31 &       1.35 &      0.110 &    11.68 &         86 &         81 &       1 \\
        36 &     SX Crv &      10.90 &       1.21 &      0.111 &    10.33 &         97 &         92 & $\surd$ \\
        37 &     VW LMi &       8.04 &       0.90 &      0.112 &     7.61 &        131 &        124 &       1 \\
        38 &     WY Hor &       8.82 &       1.00 &      0.113 &     8.34 &        120 &        113 & $\surd$ \\
        39 &     XY Leo &      15.86 &       1.80 &      0.113 &    15.00 &         67 &         63 &       1 \\
        40 &     EX Leo &       9.84 &       1.11 &      0.113 &     9.31 &        107 &        102 & $\surd$ \\
        41 &     OQ Vel &       5.37 &       0.62 &      0.115 &     5.07 &        197 &        186 & $\surd$ \\
        42 &  V1084 Sco &      11.16 &       1.32 &      0.118 &    10.50 &         95 &         90 &       1 \\
        43 &     NN Vir &       9.48 &       1.14 &      0.120 &     8.90 &        112 &        106 & $\surd$ \\
        44 &  V2377 Oph &      10.09 &       1.22 &      0.121 &     9.46 &        106 &         99 & $\surd$ \\
        45 &     VZ Psc &      16.77 &       2.07 &      0.123 &    15.68 &         64 &         60 & $\surd$ \\
        46 &   V351 Peg &       7.34 &       0.92 &      0.125 &     6.85 &        146 &        136 & $\surd$ \\
        47 &     TV Pic &       5.04 &       0.63 &      0.125 &     4.70 &        213 &        198 & $\surd$ \\
        48 &   V870 Ara &      10.01 &       1.34 &      0.134 &     9.23 &        108 &        100 & $\surd$ \\
        49 &   V839 Cen &      11.94 &       1.66 &      0.139 &    10.93 &         92 &         84 & $\surd$ \\
        50 &   V386 Pav &       7.14 &       1.01 &      0.141 &     6.52 &        153 &        140 & $\surd$ \\
        51 &     DX Tuc &       7.40 &       1.12 &      0.151 &     6.65 &        150 &        135 & $\surd$ \\
        52 &   V752 Cen &       9.51 &       1.47 &      0.155 &     8.49 &        118 &        105 & $\surd$ \\
        53 &     AQ Psc &       8.03 &       1.29 &      0.161 &     7.09 &        141 &        125 &       1 \\
        54 &     AC Boo &       7.58 &       1.27 &      0.168 &     6.60 &        152 &        132 & $\surd$ \\
        55 &     BV Dra &      14.86 &       2.56 &      0.172 &    12.82 &         78 &         67 &       1 \\
        56 &     FU Dra &       6.25 &       1.09 &      0.174 &     5.37 &        186 &        160 & $\surd$ \\
        57 &  V1073 Cyg &       5.44 &       0.95 &      0.175 &     4.66 &        214 &        184 & $\surd$ \\
\hline
\end{tabular}
}
\\
Remarks: ($\surd$): selected; (1): discarded because of membership to a multiple system.\\
\end{table}

\subsection{Other observational data} 

In addition to LK bias corrected parallaxes, the most reliable orbital 
periods and intrinsic colors are needed to revise the PLC relation.  
Orbital periods and spectral types are mostly from the catalog of
\cite{PKT03}, while the maximum visual brightnesses ($V_{max}$) 
were taken from \cite{PR06}. Intrinsic $(B-V)_{0}$ colors of main
sequence stars were taken from Neill Reid's WEB page
\footnote{http://www-int.stsci.edu/$\sim$inr/intrins.html}. In
addition to these visual photometric data, the infrared brightnesses 
$J$, $H$ and $K_{s}$ magnitudes were taken from the Point-Source
Catalogue and Atlas \citep{C03, S06} which is based on the Two
Micron All Sky Survey ({\em 2MASS}) observations. The {\em 2MASS}
photometric system comprises Johnson's $J$ (1.25 $\mu$m) and $H$
(1.65 $\mu$m) bands with the addition of $K_{s}$ (2.17 $\mu$m) band,
which is bluer than Johnson's $K$-band. Infrared data are for to 
establishing PLC relation at infrared wavelengths.

The color excess $E(B-V)$, although a relatively small quantity, it can 
effect the calibration in a systematic way. It has already been noticed by
\cite{RD97} that the color excess were likely overestimated in previous 
studies \citep{RK81, R83}. Therefore, we have carefully re-investigated the 
color excess of the stars in Table 1 using three independent methods. The 
first method uses spectral types to estimate intrinsic color of a system. 
The second method uses the color excess found from the literature
directly, which are the estimates from the hydrogen column density
or main-sequence fitting. The third method computes the color
excess from \cite{Sc98} maps by using NASA Extragalactic
Database\footnote{http://nedwww.ipac.caltech.edu/forms/calculator.html}.
Since sample W UMa stars are relatively nearby, the color excesses
according to \cite{Sc98} need to be reduced. First, $E_{\infty}(B-V)$ 
color excess in the galactic latitude ($b$) and longitude ($l$) was 
taken from \cite{Sc98}. Then, the total absorption towards the star in 
the galactic disk in the $V$-band was evaluated as
\begin{equation}
A_{\infty}(b)=3.1E_{\infty}(B-V),
\end{equation}
where the subscript symbolizes up to infinity but actually it is up to 
the edge of our galaxy in the line of sight. Thirdly, the interstellar 
absorption up to the star distance $d$ is calculated according to \cite{BS80}
\begin{equation}
A_{d}(b)=A_{\infty}(b)\Biggl[1-exp\Biggl(\frac{-\mid d~sin(b)\mid}{H}\Biggr)\Biggr],
\end{equation}
where $H$ is the scaleheight for the interstellar dust which is adopted to 
be 100 pc as usual \citep[see e.g.][]{Mendez98}. Finally, the color excess for 
the star at the distance $d$ is estimated as
\begin{equation}
E_{d}(B-V)=A_{d}(b)~/~3.1.
\end{equation}
It is demonstrated (Fig. 2) that the first method usually overestimates the color 
excess with respect to the other two methods. The color excesses from the literature
and the values computed using \cite{Sc98} extinction maps agree with
each other. Literature values could not be trusted especially if they are zero, 
which could be the cases because the color excess was ignored blindly. On the
other hand, values computed by the method of \cite{Sc98} extinction
maps are not trusted at low galactic latitudes ($|b|<10^{\circ}$)
due to local inhomogeneities of galactic disc. Therefore, care was 
given to $E(B-V)$ values of all stars one by one and the most 
trustable value was selected. Adopted values are usually the ones which
are taken from the literature where the source references are given
or the method of inquiry is indicated (Table 2).

\begin{figure}
\center
\resizebox{80mm}{42.7mm}{\includegraphics*{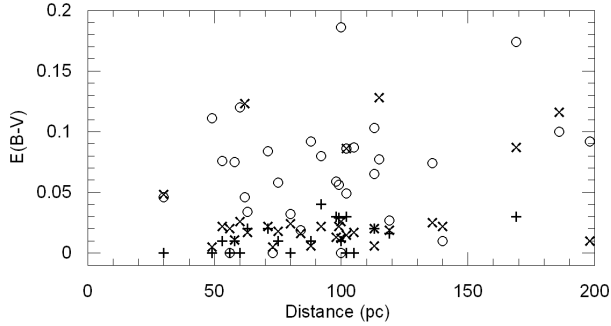}}
\caption{Color excesses; (o) observed minus intrinsic, ($+$) literature and ($\times$) Schlegel et al. (1998) reduction.}
\end{figure}

Once color excesses are known, then the intrinsic colors and the interstellar absorption in the $V$-band were computed as
\begin{equation}
(B-V)_{0}=(B-V)-E(B-V),
\end{equation}
\begin{equation}
A_{V}=3.1E(B-V).
\end{equation}
Consequently, de--reddening of the infrared bands becomes ready using the relations given by \cite{FM03},
\citep [see also,][]{B06, Ak07,B08}, which are
\begin{equation}
J_{0}=J-0.887E(B-V),
\end{equation}
\begin{equation}
(J-H)_{0}=(J-H)-0.322E(B-V),
\end{equation}
\begin{equation}
(H-K_{s})_{0}=(H-K_{s})-0.183E(B-V).
\end{equation}

\subsection{Formation of the final calibration list}
Regarding the accuracy, today's technology is capable of determining 
a visual brightness within the accuracy of few millimagnitudes.
Because of O'Connell's effect, which could be several millimagnitudes 
\citep{D84}, the maximum visual brightness ($V_{max}$) of W UMa stars 
have additional uncertainties. Error contribution of a visual 
brightness, even if it is in the order of several millimagnitudes, is 
negligible compared to the other contributing errors. For example, 
uncertainties of $E(B-V)$, which contributes to the uncertainty of 
interstellar absorption through $A_{V}=3.1E(B-V)$, are in the order 
of 0.01 -- 0.02 mag \citep{RD97, R02, R06}. The uncertainty
of $A_{V}$, however, is not only due to the uncertainty of $E(B-V)$, 
but also due to the uncertainty of the coefficient in front of $E(B-V)$. 
Depending upon the position of a star relative to the Sun in the galactic
plane, there could be cases where the coefficient could be as small as 2.75 
and as big as 3.52 \citep{CCM89}. Therefore a rough estimate of the 
uncertainty of $A_{V}$  is in average 0.06 magnitudes. Even if the error
contribution of $A_{V}$ is overestimated to be 0.1 mag, it would still be 
smaller than the uncertainty contribution of a 10\% parallax error, 
which is 0.217 mag.

There are 29 W UMa systems with $\sigma_{\pi}/\pi < 0.10$ in Table 1, 
which could be considered statistically significant at least at the limit.
However, \cite{PR06} have found that up to 59\% of W UMa stars have 
companions. Actually, one of the criteria of multiplicity for \cite{PR06}
was the large error of the observed parallax. Trigonometric parallaxes are 
frequently wrong for these systems because complexity of the multiple 
orbital motions shows itself both at proper motion and parallax measurements.
Therefore, W UMa binaries which are known to be associated with companions 
were excluded as \cite{R06} did when establishing PL relation.
W UMa which are known to have companions are marked in Table 1. The 13 out 29 stars 
with $\sigma_{\pi}/\pi<0.10$ are marked, so after removing them, the 16 
W UMa systems left would definitely not satisfy statistical significance. We decided 
to include all systems up to the limit $\sigma_{\pi}/\pi<0.15$, which means
uncertainties in the absolute magnitudes would be better than 0.33 mag. This process 
provided us with 31 W UMa stars after removing the ones with companions, which we think 
satisfy statistical significance and accuracy for re-calibrating the PLC relation.
W UMa stars with larger parallax errors are not included to avoid adding stars with 
less accurate absolute magnitudes.    

In Table 2, LK corrected absolute magnitudes $M_{V}$ and $M_{J}$ for 31 W UMa 
stars were computed with most probable true parallax and adopted color excess 
$E(B-V)$. The uncorrected absolute magnitudes [$M_{V}(Hip)$ and $M_{J}(Hip)$] 
were also listed just for comparison. Fig. 3 compares corrected and uncorrected 
distances of the calibration sample.

\begin{table*}
\begin{center}
\setlength{\tabcolsep}{3pt} 
\begin{minipage}{\textwidth}
\caption{Calibration sample and data used in recalibration of the PLC relation for W UMa stars.}
{\scriptsize
\begin{tabular}{lcccccccccccccccccc}
\hline
(1) & (2) & (3) &  (4) & (5) & (6)& (7) & (8) & (9) & (10) & (11) & (12) & (13) & (14) & (15) & (16) & (17) & (18) & (19) \\
Name & $\pi $ & $\sigma_{\pi}/{\pi}$ &  $P$ & $SpT$ & (B-V)$_{0}$ & E(B-V) & E(B-V) & E(B-V) &  $V_{max}$ & (B-V) & E(B-V)  &  J & (J-H) & (H-K$_{s}$) & M(V) & M(V) & M(J) & M(J)   \\
     & (mas)  &                      & (day)&      &  INT$^{1}$ & OMI$^{2}$ & LIT$^{3}$  & SCH$^{4}$ & (mag) & OBS$^{5}$  & ADP$^{6}$  & (mag)  & (mag) & (mag)     &  (Hip)  &   (true)     &  (Hip)    &  (true)            \\
\hline
$\epsilon$ CrA&33.32&0.028&0.59144070 & F2V    & 0.35 & 0.046 & 0.000 & 0.048 & 4.74 & 0.396 & 0.046$^{~a}$ & 4.052 & 0.264 &-0.090 & 2.218 & 2.211 & 1.632 & 1.625\\
V972 Her & 16.16 & 0.038 & 0.44309740 & F4V    & 0.42 & 0.046 &  --   & 0.123 & 6.62 & 0.466 & 0.046$^{~a}$ & 5.785 & 0.149 & 0.103 & 2.532 & 2.519 & 1.798 & 1.786\\
AE Phe   & 20.36 & 0.040 & 0.36237274 & F8V    & 0.53 & 0.111 & 0.000 & 0.005 & 7.56 & 0.641 & 0.000$^{~b}$ & 6.577 & 0.289 & 0.107 & 4.118 & 4.104 & 3.135 & 3.121\\
DO Cha   & 12.46 & 0.045 & 0.68144600 & F7V    & 0.50 & 0.032 & 0.000 & 0.024 & 7.61 & 0.535 & 0.032$^{~a}$ & 6.634 & 0.250 & 0.067 & 3.006 & 2.988 & 2.101 & 2.083\\
S Ant    & 13.15 & 0.053 & 0.64834550 & A9V:   & 0.30 & 0.058 & 0.010 & 0.018 & 6.40 & 0.358 & 0.010$^{~b}$ & 6.087 & 0.127 & 0.095 & 1.988 & 1.963 & 1.697 & 1.673\\
OU Ser   & 17.10 & 0.055 & 0.29676450 & F9/G0V & 0.56 & 0.075 & 0.010 & 0.011 & 8.10 & 0.635 & 0.010$^{~b}$ & 6.946 & 0.264 & 0.112 & 4.260 & 4.234 & 3.129 & 3.102\\
V759 Cen & 15.66 & 0.059 & 0.39399912 & F9V    & 0.56 & 0.034 & 0.020 & 0.017 & 7.40 & 0.594 & 0.020$^{~b}$ & 6.530 & 0.242 & 0.063 & 3.342 & 3.311 & 2.517 & 2.486\\
GR Vir   & 18.53 & 0.063 & 0.34696950 & F7-8V  & 0.50 & 0.076 & 0.010 & 0.022 & 7.80 & 0.576 & 0.010$^{~b}$ & 7.048 & 0.242 & 0.044 & 4.143 & 4.108 & 3.413 & 3.378\\
YY Eri   & 17.63 & 0.067 & 0.32150003 & G5V    & 0.68 &-0.006 & 0.000 & 0.020 & 8.10 & 0.674 & 0.000$^{~b}$ & 7.031 & 0.337 & 0.106 & 4.372 & 4.331 & 3.303 & 3.262\\
IS CMa   &  9.80 & 0.071 & 0.61698000 & F3V    & 0.38 &-0.008 & 0.010 & 0.012 & 6.96 & 0.372 & 0.010$^{~b}$ & 6.619 & 0.151 & 0.071 & 1.931 & 1.886 & 1.612 & 1.567\\
CP Hyi   &  8.20 & 0.072 & 0.47940600 & F0V    & 0.32 & 0.027 & 0.016 & 0.019 & 7.80 & 0.347 & 0.016$^{~c}$ & 7.187 & 0.105 & 0.099 & 2.367 & 2.320 & 1.789 & 1.743\\
YY CrB   & 11.10 & 0.075 & 0.37656400 & F8V    & 0.53 & 0.092 & 0.010 & 0.006 & 8.48 & 0.622 & 0.010$^{~b}$ & 7.640 & 0.234 & 0.097 & 3.726 & 3.675 & 2.908 & 2.857\\
V757 Cen & 13.83 & 0.078 & 0.34316916 & F9V    & 0.56 & 0.084 & 0.020 & 0.022 & 8.30 & 0.644 & 0.020$^{~b}$ & 7.336 & 0.304 & 0.109 & 3.996 & 3.941 & 3.077 & 3.022\\
FX Eri   & 13.33 & 0.078 & 0.29234500 & G9:    & 0.78 &-0.007 &   --  & 0.005 & 9.56 & 0.773 & 0.000$^{~a}$ & 8.152 & 0.407 & 0.108 & 5.239 & 5.184 & 3.831 & 3.776\\
RR Cen   &  9.46 & 0.087 & 0.60569200 & F0V    & 0.32 & 0.086 & 0.030 & 0.086 & 7.27 & 0.406 & 0.086$^{~a}$ & 6.765 & 0.113 & 0.078 & 1.951 & 1.882 & 1.636 & 1.567\\
GM Dra   &  9.84 & 0.087 & 0.33874120 & F5V    & 0.45 & 0.059 & 0.030 & 0.013 & 8.66 & 0.509 & 0.030$^{~b}$ & 7.736 & 0.252 & 0.043 & 3.601 & 3.533 & 2.744 & 2.675\\
V535 Ara &  8.49 & 0.101 & 0.62930107 & A8V:   & 0.27 & 0.065 & 0.020 & 0.020 & 7.17 & 0.335 & 0.020$^{~b}$ & 6.989 & 0.149 & 0.021 & 1.848 & 1.753 & 1.711 & 1.616\\
TY Men   &  5.65 & 0.106 & 0.46166680 & A3/5V  & 0.08 & 0.174 & 0.030 & 0.087 & 8.08 & 0.254 & 0.104$^{~a}$ & 7.597 & 0.088 & 0.075 & 1.623 & 1.518 & 1.370 & 1.265\\
V918 Her &  8.28 & 0.107 & 0.57481000 & A7V    & 0.20 & 0.077 &   --  & 0.128 & 7.31 & 0.277 & 0.077$^{~a}$ & 6.703 & 0.095 & 0.013 & 1,769 & 1.662 & 1.332 & 1.225\\
SX Crv   & 10.33 & 0.111 & 0.31662090 & F6V    & 0.48 & 0.080 & 0.040 & 0.022 & 8.99 & 0.560 & 0.040$^{~b}$ & 7.927 & 0.267 & 0.075 & 4.053 & 3.937 & 3.079 & 2.963\\
WY Hor   &  8.34 & 0.113 & 0.39894000 & G2IV/V & 0.63 & 0.103 &  --   & 0.006 & 9.39 & 0.733 & 0.103$^{~a}$ & 8.335 & 0.321 & 0.105 & 3.798 & 3.678 & 2.971 & 2.850\\
EX Leo   &  9.31 & 0.113 & 0.40860250 & F6V    & 0.48 & 0.049 & 0.000 & 0.015 & 8.13 & 0.529 & 0.000$^{~b}$ & 7.332 & 0.217 & 0.068 & 3.095 & 2.974 & 2.297 & 2.176\\
OQ Vel   &  5.07 & 0.115 & 0.58133800 & A3IV   & 0.08 & 0.100 &   --  & 0.116 & 7.67 & 0.180 & 0.100$^{~a}$ & 7.323 & 0.037 & 0.068 & 1.010 & 0.885 & 0.884 & 0.759\\
NN Vir   &  8.90 & 0.120 & 0.48071484 & F0/1V  & 0.32 & 0.087 & 0.000 & 0.017 & 7.60 & 0.407 & 0.000$^{~c}$ & 7.060 & 0.126 & 0.114 & 2.484 & 2.347 & 1.944 & 1.807\\
V2377 Oph&  9.46 & 0.121 & 0.42540100 & G0/1V  & 0.60 & 0.056 & 0.029 & 0.022 & 8.45 & 0.656 & 0.029$^{~c}$ & 7.350 & 0.275 & 0.058 & 3.380 & 3.239 & 2.344 & 2.204\\
VZ Psc   & 15.68 & 0.123 & 0.26125918 & K5V    & 1.15 & 0.120 & 0.000 & 0.026 &10.15 & 1.270 & 0.000$^{~d}$ & 8.158 & 0.576 & 0.176 & 6.273 & 6.127 & 4.281 & 4.135\\
V351 Peg &  6.85 & 0.125 & 0.59329700 & A8V    & 0.27 & 0.074 &   --  & 0.025 & 7.92 & 0.344 & 0.074$^{~a}$ & 7.266 & 0.099 & 0.038 & 2.019 & 1.869 & 1.529 & 1.378\\
TV Pic   &  4.70 & 0.125 & 0.85198700 & A2V    & 0.05 & 0.092 &   --  & 0.010 & 7.37 & 0.142 & 0.092$^{~d}$ & 7.096 & 0.027 & 0.072 & 0.597 & 0.446 & 0.527 & 0.376\\
V870 Ara &  9.23 & 0.134 & 0.39978000 & F8     & 0.53 & 0.186 &   --  & 0.026 & 8.80 & 0.716 & 0.026$^{~e}$ & 7.741 & 0.264 & 0.089 & 3.722 & 3.545 & 2.720 & 2.544\\
V839 Cen & 10.93 & 0.139 & 0.33093400 & G2:    & 0.63 & 0.019 &   --  & 0.016 & 8.82 & 0.649 & 0.019$^{~a}$ & 8.315 & 0.268 & 0.146 & 4.146 & 3.955 & 3.683 & 3.492\\
V386 Pav &  6.52 & 0.141 & 0.55184100 & A9V    & 0.30 & 0.010 &   --  & 0.022 & 8.28 & 0.310 & 0.010$^{~a}$ & 7.759 & 0.101 & 0.091 & 2.517 & 2.320 & 2.019 & 1.821\\
\hline
\end{tabular}\\
{1: intrinsic color, 2: Observed minus intrinsic colors, 3: taken from literature, 4: based on Schlegel et al. (1998), 5: Observed color, 6: Adopted color excess \\
 a: Color excess of observed minus intrinsic colors, b: \cite{R06}, c: \cite{Nordstrom04}, d: \cite{RD97}, e: \cite{Sc98} }
}
\end{minipage}
\end{center}
\end{table*}

\begin{figure}
\begin{center}
\resizebox{70mm}{70mm}{\includegraphics*{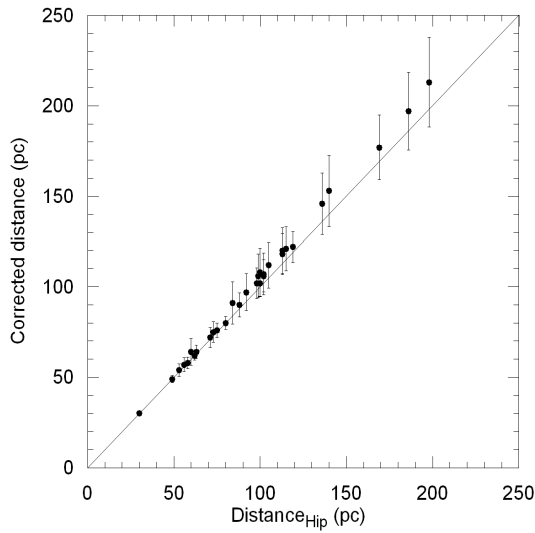}} 
\end{center}
\caption{Comparing corrected and uncorrected ({\em Hipparcos}) distances of W UMa stars 
with parallax errors $\sigma_{\pi}/\pi<0.15$.}
\end{figure}

\section{Calibrations and discussions}
\subsection{PLC relation according to uncorrected {\em Hipparcos} parallaxes}

There were 40 contact binaries used by \cite{RD97} when establishing 
former PLC relation. Only 11 of them ($\epsilon$ CrA, AE Phe, V759 Cen, 
V757 Cen, RR Cen, GR Vir, YY Eri, V535 Ara, TY Men, SX Crv, VZ Psc) are 
retained in our final calibration sample (Table 2). The others are eliminated 
either because they are members of multiple systems or their parallax errors 
are out of the acceptable limits of this study.

In the first step, we were curious if the coefficients of the PLC 
relation by \cite{RD97} would change. Therefore, the uncorrected $M_{V}$ of 
the W UMa stars in the calibration list are used in a regression analysis, 
which uses least squares fit for determining the three coefficients 
appearing in the PLC relation. The regression analysis produced similar 
coefficients. Therefore, the PLC relation of \cite{RD97} were kept as 
the reference relation with LK bias (uncorrected).

Fig. 4a displays the absolute visual magnitudes which are predicted 
by the PLC relation of \cite{RD97} against the absolute visual 
magnitudes from the {\em Hipparcos} parallaxes directly. According 
to the regression analysis, the correlation coefficient $R^{2}=0.916$ 
and the standard deviation from the diagonal is $s=0.35$.

So, it can be concluded that having only more accurate parallaxes ($\sigma_{\pi}/\pi<0.15$)
than the original sample ($\sigma_{\pi}/\pi<0.24$) of \cite{RD97}, the result does  not justify 
re-calibrating the PLC relation, but only confirms an existing relation.

\begin{figure}
\center
\resizebox{80mm}{157.3mm}{\includegraphics*{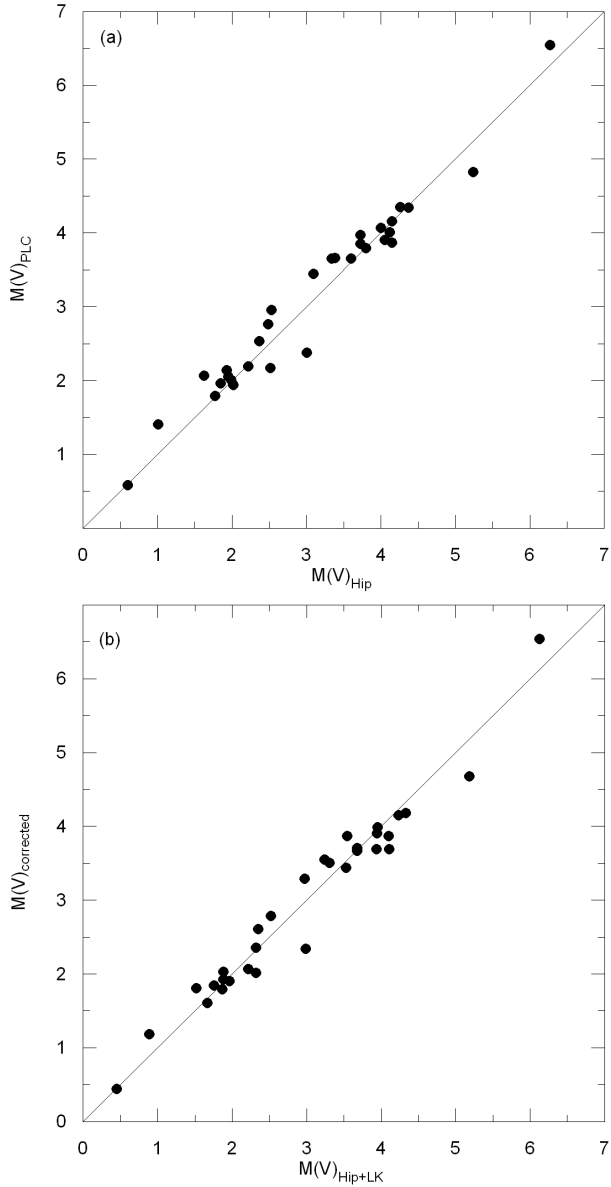}}
\caption{a) Absolute visual magnitudes predicted by the PLC relation of Rucinski \& Duerbeck (1997)
versus absolute visual magnitudes of {\em Hipparcos} parallaxes. b) absolute visual magnitudes predicted
by PLC relation corrected for LK bias versus absolute visual magnitudes corrected for LK bias.}
\end{figure}

\subsection{PLC relation according to LK bias corrected parallaxes}
Using the corrected absolute magnitudes ($M_{V}$), orbital periods ($P$) and the 
intrinsic $(B-V)_{0}$ colors determined from the observed colors and adopted color 
excesses ($E(B-V)$), of the stars in Table 2, the coefficients of the linear equation
\begin{equation}
M(V)_{NEW}=a\log P + b(B-V)_{0}+ c,
\end{equation}
have been determined as $a=-3.88(\pm0.62)$, $b=3.36(\pm0.34)$ and
$c=0.0044(\pm 0.14)$ by a regression analysis with a correlation
coefficient $R^{2}=0.959$ and standard deviation $s=0.27$. According
to the regression analysis it can be  concluded that the internal
error of the new relation is $s.e.=s/\sqrt{N}=0.27/\sqrt{31}=0.048$
mag. That is, in a case if periods and intrinsic colors are 
errorless, then the standard error of predicted $M_{V}$ is expected to be $\mp
0.048$ mag. The standard error of $M_{V}$ should become larger if errors 
are introduced through orbital periods and intrinsic colors appearing in 
the relation.

\subsection{Comparing PLC relations with and without LK bias}
Comparing PLC relations of same format with and without LK bias may help to analyze and study
the effect of LK bias on the PLC relation. There could be two basic types of comparison:
The two relations may be compared on their own grounds to see how good their predicting powers 
are according to their own conditions; or they could be compared to see what the 
difference is in predicting true absolute magnitudes. 

Consequently, for the first type of comparison, the \\
$M(V)_{NEW}$ points are plotted against the corrected absolute magnitudes coming from the 
corrected parallaxes in Fig. 4b. Although, the correlation coefficient is slightly better 
($R^{2}=0.928$) and the standard deviation $s=0.32$ is smaller, the apparency of Fig. 4b is not 
much different Fig. 4a. Both figures are consistent with each other. Thus, LK correction is 
justified. 

The difference between the two PLC relations can be observed better in Fig. 5
where the predictions of both relations are plotted against the corrected
absolute magnitudes. Although both data sets remain within the $2\sigma$ reliability limit, 
it is clearly shown in Fig. 5 that the PLC relation of \cite{RD97} systematically 
underestimates the absolute brightness of the same sample stars.

\begin{figure}
\begin{center}
\resizebox{80mm}{96.7mm}{\includegraphics*{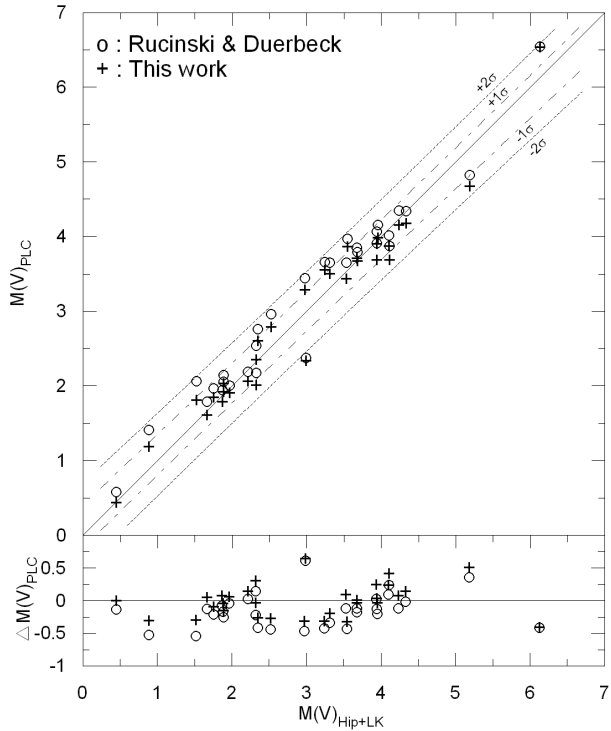}}
\end{center}
\caption{Absolute visual magnitudes predicted by two PLC relations (Rucinski \& Duerbeck
1997 and this work) plotted against the absolute visual magnitudes directly from Pogson's
relation using LK bias corrected parallaxes. The dashed-dotted and dashed lines represent
prediction limits for $\pm 1\sigma$ and $\pm 2\sigma$, respectively.}
\end{figure}
\begin{figure}
\center
\resizebox{80mm}{96.7mm}{\includegraphics*{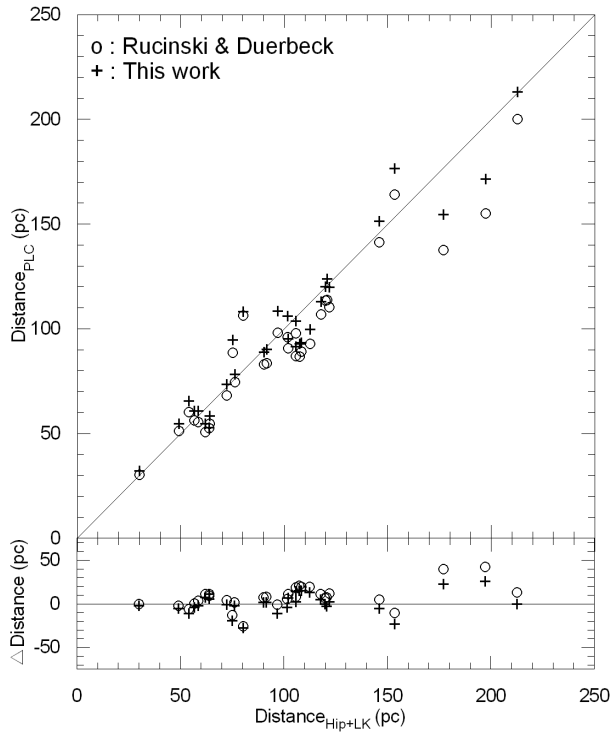}}
\caption{The distances according to both PLC relations are plotted against the LK bias 
corrected distances.}
\end{figure}

A similar comparison can be done on a distance scale rather than a
magnitude scale. Fig. 6 displays such a comparison. It is clearly
shown that the PLC relation of \cite{RD97} underestimates the
distances with respect to the predictions of the PLC relation from
this study especially when the distances are large. If the distances
are smaller than 100 pc, apparently, the difference becomes
negligible. This must be because; LK corrections become noticeable
if $\sigma_{\pi}/\pi>0.05$ and increases as $\sigma_{\pi}/\pi$ increases. 

\subsection{Calibrating PLC relation using {\em 2MASS} data}
PLC relation of W UMa stars could be calibrated using different color indices and color
excesses. A PLC relation using $(V-I)_{0}$ already exists \citep{R04}. The advantage
of shifting PLC relation towards the infrared colors is obvious; extinction is less, so
uncertainty of interstellar absorption becomes relatively reduced. 

The method of calibration is the same. Absolute magnitudes, and then intrinsic colors through
(8) and (9) were computed using the data in Table 2 for the same sample (31 stars). Finally
a regression analysis is done for determining the numerical coefficients. It has been tried 
for many different forms of the relation such as; with one color or both $(J-H)_{0}$ and 
$(H-K_{s})_{0}$ and even to include $(B-V)_{0}$ color besides the terms of $\log P$ including a 
free constant. The highest correlation have been found with the following
\begin{equation}
M(J) =a\log P + b(J-H)_{0}+ c(H-K_{s})_{0}+d,
\end{equation}
where $a=-3.17(\pm0.62)$, $b=3.84(\pm0.62)$, $c=1.81(\pm1.16)$ and $d=0.24(\pm 0.14)$, which gives
a correlation coefficient $R^{2}=0.927$ and a standard deviation $s=0.26$ mag.
Consequently, the internal error of the relation is $\mp0.047$ mag. Since best
correlation was achieved by using two rather than a single color term, it seems
using two color terms increases the correlation. This is expected because each color 
term contributes to improving the correlation. The constant term which has an absolute value 
comparable to its uncertainty may not be real so it can be avoided in the relation.  

The predicted $M(J)$ values by Eq. 11 are plotted against LK corrected absolute magnitudes (Fig. 7). 
Despite, {\em 2MASS} data are single epoch observations, which were not guaranteed to
be at the maximum brightness phase of the systems, the calibration gave 
reliable relation which can be used in predicting the true distances of W UMa stars
from their periods and infrared colors. It is even more interesting that the standard
deviation of data and the internal error of the PLC relation in infrared are slightly
smaller than the standard deviation and internal error of the PLC relation in the visual
although the PLC relation is slightly less correlated at infrared.

The distance predictions of the PLC relation are compared to LK bias corrected distances in 
Fig. 8. The internal error of the relation is $\mp0.022$. The PLC relation at infrared could 
have been improved more if maximum brightnesses at these infrared colors were  available.

\begin{figure}
\begin{center}
\resizebox{80mm}{96.7mm}{\includegraphics*{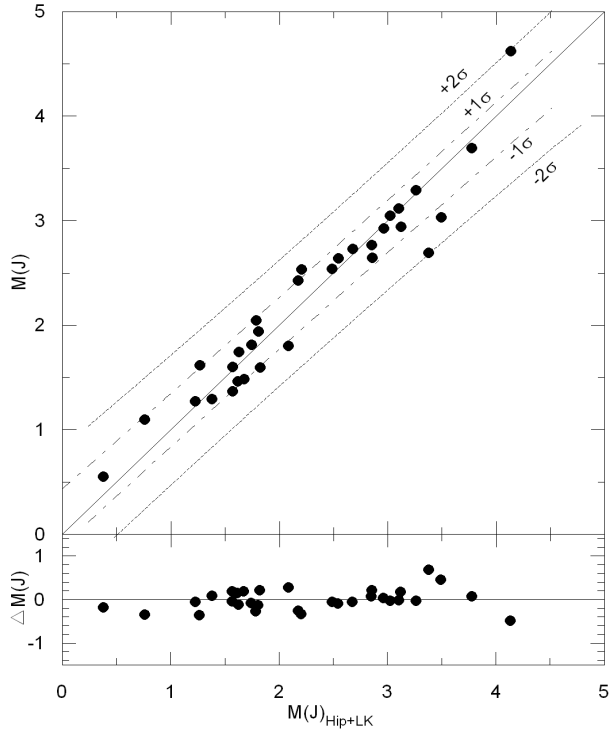}}
\end{center}
\caption{Predicted $M(J)$ are compared to LK corrected absolute magnitude. The dashed-dotted and 
dashed lines represent prediction limits for $\pm 1\sigma$ and $\pm 2\sigma$, respectively.}
\end{figure}

\begin{figure}
\center
\resizebox{80mm}{96.7mm}{\includegraphics*{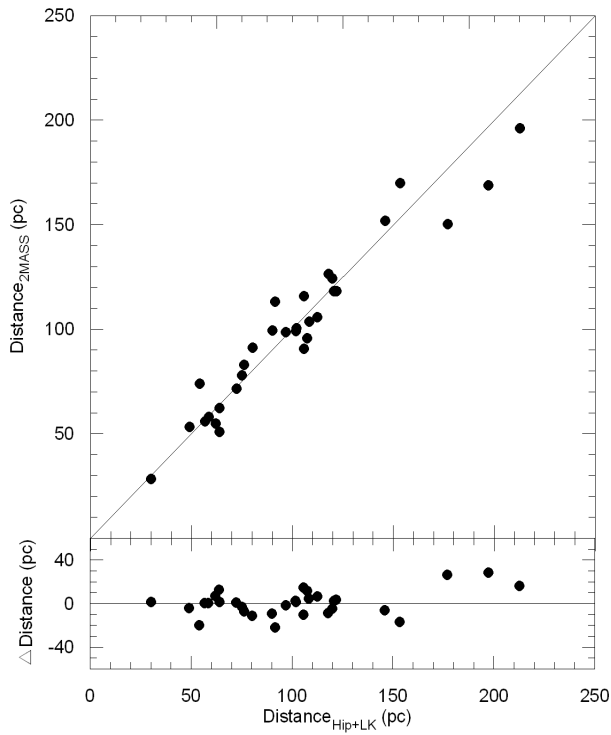}}
\caption{Distances predicted by the PLC relation in the infrared are compared to the LK 
corrected distances.}
\end{figure}

\section{Conclusions}
In this study, a statistical bias, which is classically known as Lutz-Kelker (LK) 
bias, has been introduced to the PLC relation of W UMa stars. Since parallax errors 
remain after a LK bias correction, studies such as ours, should still
be careful in setting up an upper limit on the reliability of input parallax data. 
In this study, we have preferred to work with parallax data more accurate than 15\%.

Using LK corrected parallaxes (or distances), we have re-calibrated the classical
PLC relation of \cite{RD97}. The new relation has been compared with the classical 
relation. It has been found that old and new relations are nearly equivalent in 
predicting a distance up to about 100 parsecs. The difference between the predictions 
increases and may become intolerable as the distances of stars increase. The new 
relation is valid in the ranges ($0.26< P(day)< 0.87$, $0.04<(B-V)_{0}<1.28$ and 
$0.43<M(V)<6.55$).

We have also produced a PLC relation which can be used by {\em 2MASS} data.
Despite {\em 2MASS} data are {\it single epoch} observations, which are not
guaranteed to be at the maximum brightness phase of the W UMa stars in the calibration sample
(Table 2), the established relation has been found surprisingly consistent and
reliable to predict LK corrected distances. This new relation is valid in the
ranges ($0.26< P(day)< 0.87$, $-0.01<(J-H)_{0}<0.58$, $-0.10<(H-K_{s})_{0}<0.18$
and $0.54<M(J)<4.63$). Observers are encouraged to obtain light and color curves of
W UMa stars at {\em 2MASS} colors. If the apparent magnitudes at maximum brightness
phases of W UMa stars are provided with {\em 2MASS} colors, the current PLC relation
using infrared colors could be improved to provide more reliable distances.

\section{Acknowledgments}
This work has been partially supported by T\"UB\.ITAK 104T508 and 106T688. We would also 
like to thank Prof. Dr. Utku Co\c skuno\u glu and Ba\c sar Co\c skuno\u glu correcting the 
english grammer and linguistics of the manuscript. This research has made use of the SIMBAD 
database, operated at CDS, Strasbourg, France. This publication makes use of data 
products from the Two Micron All Sky Survey, which is a joint project of the University 
of Massachusetts and the Infrared Processing and Analysis Center/California Institute 
of Technology, funded by the National Aeronautics and Space Administration and the 
National Science Foundation. This research has made use of the NASA/IPAC Extragalactic 
Database (NED) which is operated by the Jet Propulsion Laboratory, California Institute 
of Technology, under contract with the National Aeronautics and Space Administration.

\end{document}